\documentclass[final]{vgtc}                          % final (conference style)

\usepackage{microtype}  
\usepackage{graphicx}
\usepackage{textcomp}               
\usepackage{caption}
\usepackage{mathptmx}                 
\usepackage{times}                    
\usepackage{cite}                     
\usepackage{tabu}  
\usepackage{booktabs}                
\usepackage{enumitem}
\usepackage{mathptmx}
\usepackage{xargs}                      
\usepackage[pdftex]{xcolor}

\onlineid{0}

\vgtccategory{Workshop}

\vgtcinsertpkg

\title{Improving X-ray Diagnostics through Eye-Tracking and XR}

\author{Catarina Moreira\thanks{e-mail: catarina.pintomoreira@qut.edu.au}\\ 
        \scriptsize INESC-ID / School of Information Systems, \\
      \scriptsize Queensland University of Technology, Australia \\
    \and Isabel Blanco Nobre\thanks{e-mail: isabel.blanco.nobre@lusiadas.pt}\\
       \scriptsize  Imagiology Department\\ 
        \scriptsize  Grupo Lusíadas, Lisboa, Portugal\\
    \and Sandra Costa Sousa\thanks{e-mail: sandra.costa.sousa@lusiadas.pt} \\
 \scriptsize Imagiology Department\\
  \scriptsize  Grupo Lusíadas, Lisboa, Portugal \\
  \and Jo\~{a}o Madeiras Pereira\thanks{e-mail: jap@inesc-id.pt}\\ 
     \scriptsize  INESC-ID / Instituto Superior T\'{e}cnico \\
      \scriptsize Universidade de Lisboa, Portugal \\
     \and Joaquim Jorge\thanks{e-mail: jorgej@tecnico.ulisboa.pt}\\ 
     \scriptsize INESC-ID / Instituto Superior T\'{e}cnico \\
     \scriptsize Universidade de Lisboa, Portugal
  }

\abstract{There is a growing need to assist radiologists in performing X-ray readings and diagnoses fast, comfortably, and effectively. As radiologists strive to maximize productivity, it is essential to consider the impact of reading rooms in interpreting complex examinations and ensure that higher volume and reporting speeds do not compromise patient outcomes. Virtual Reality (VR) is a disruptive technology for clinical practice in assessing X-ray images. We argue that conjugating eye-tracking with VR devices and Machine Learning may overcome obstacles posed by inadequate ergonomic postures and poor room conditions that often cause erroneous diagnostics when professionals examine digital images.}

\CCScatlist{
\CCScatTwelve{Applied Computing}{Life and medical Sciences}{Health informatics}{}
\CCScatTwelve{Computing methodologies}{Artificial intelligence}{Knowledge representation and reasoning}{Causal reasoning and diagnostics}
\CCScatTwelve{Human-centered computing}{Virtual reality}{}{}
}

\begin{document}

\firstsection{Introduction}

\maketitle

The outbreak of recent pandemics affecting patients' lungs has reminded the scientific community that there is a growing need to assist radiologists in performing X-ray readings and diagnoses fast, comfortably, and effectively. As radiologists strive to maximize productivity, it is essential to consider the impact of radiologists' reading rooms in the interpretation of increasingly complex examinations and ensure that higher volume and reporting speeds do not compromise patient outcomes \cite{Waite17}. In this pressured setting, inadequate ergonomic postures and improper room conditions can cause erroneous diagnostics when professionals examine 3D digital images using common 2D desktop displays \cite{Samei05}.

Current radiology reading rooms still face many limitations. The most significant is that the quality of the X-ray images is seriously affected by the quality of the display devices, which impacts the radiologists reading ability. Additionally, the room's improper light and luminance conditions together with pressures in radiologists to increase productivity can lead to visual tiredness, cognitive overload, and decision fatigue \cite{Sokolovskaya15,Reiner12}, which makes the radiologists' readings more susceptible to biases and errors \cite{Saposnik16biases}.

Recently, Sousa et al. \cite{sousa17vrrroom, paulo-2022} showed that the application of Virtual Reality (VR) technologies for radiology reading rooms constitutes a viable, flexible, portable, and cost-effective option to address the limitations associated with traditional settings. However, one significant aspect that the authors have not explored is the potential of using radiologists' eye-tracking data to build decision models that could support, amplify, augment and enhance expert radiologists' performance when assessing X-ray images. 

\begin{figure}[!t]
     \resizebox{\columnwidth}{!} {
    \includegraphics{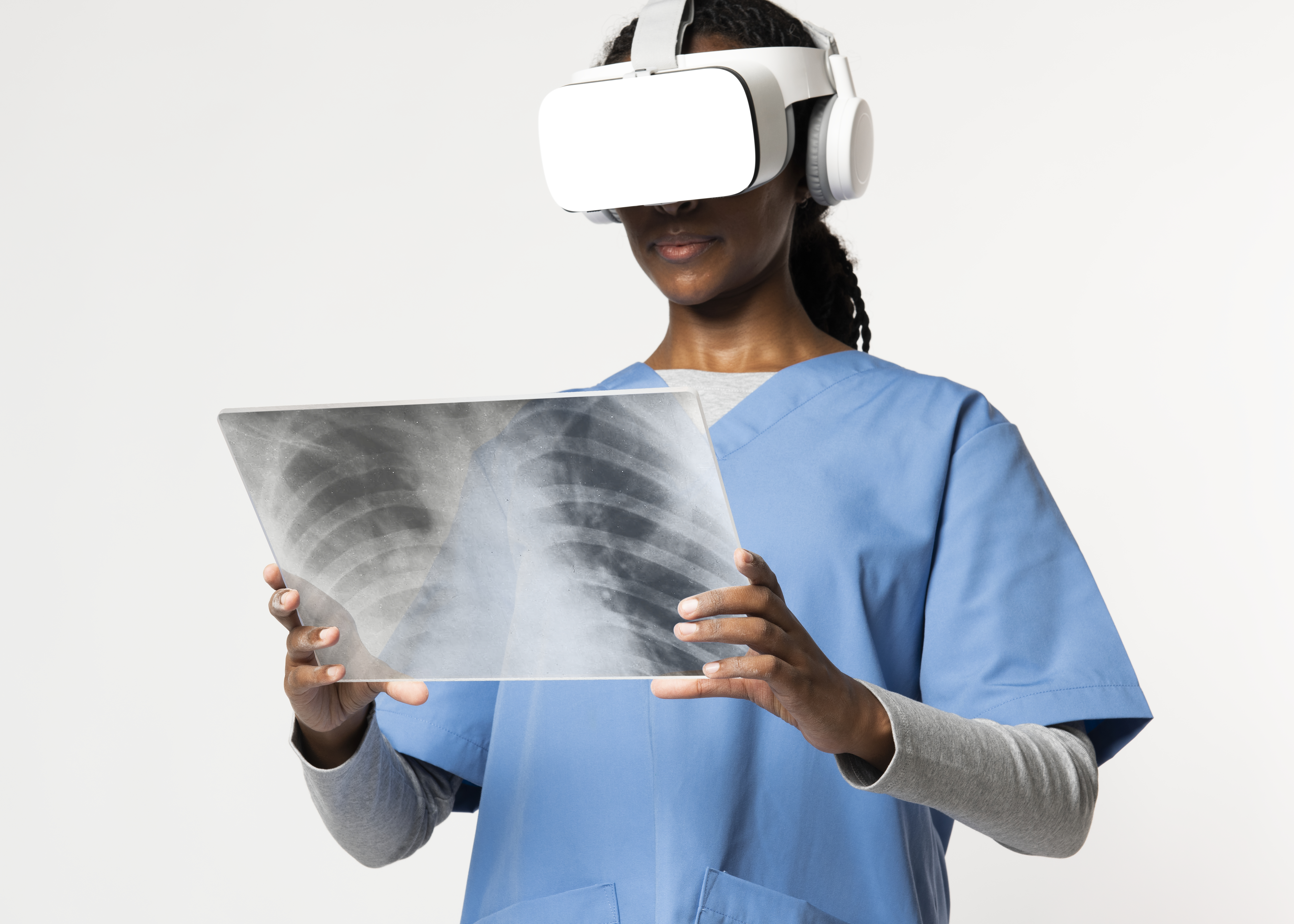}
    }
    \caption{Using VR for XRay Diagnostics. \small{Courtesy of rawpixel.com  (\url{https://www.freepik.com/photos/technology})}}
    \label{fig:xray_vr}
\end{figure}

\subsection{The Importance of Eye Tracking in Radiology}

Eye-tracking data in X-ray images have been receiving much interest in the scientific community, particularly in the field of Human-Computer Interaction (HCI) and Artificial Intelligence (AI). Although there is a rich literature correlating eye-tracking data with user cognitive load \cite{Harezlak2018eye-tracking, Borys2017}, the connection between eye movements and the cognitive state has mainly been under-exploited in Human-Computer Interaction, and a deeper understanding of its usability in practice is still required \cite{Wang21}.

In AI, eye tracking fixation data could provide an automated approach to label large amounts of data without asking expert radiologists to manually annotate them \cite{Lanfredi2021reflacx}. Although AI cannot replace clinicians in medical diagnosis, it can support expert radiologists in time-consuming tasks, such as examining chest X-rays for signs of pneumonia. AI models can outperform humans  at specific diagnoses in X-ray images. However, many times these models get the predictions correct, but not for the right clinical reasons \cite{Saporta2021,Richens2020}. Consequently, radiologists are reluctant to blindly accept the answers produced by AI without understanding how those predictions came to be. This uncertainty poses a significant barrier to adopting AI-based technologies in healthcare because they are highly susceptible to well-documented biases (such as automation bias), leading to a misdiagnosis that may erode trust in model predictions, limiting clinical translation and adoption. This has led to new research paths, such as eXplainable Artificial Intelligence (XAI), which aims to allow human users to understand and trust the results and outputs of machine learning algorithms \cite{Holzinger19}.

One of the main points that we argue in this paper is that eye-tracking data can enable novel AI architectures that, instead of learning solely from X-ray pixel data, will use radiologists' classification patterns encoded in their eye movements. Ultimately, eye-tracking data can be instrumental in teaching machines how humans learn and using that knowledge to help clinicians better understand how machines make decisions. However, current (desktop) eye-tracking technologies still suffer from significant problems that prevent the widespread collection of high-quality ocular data.

\subsection{Limitations of Desktop Eye-Tracking Devices}

Related works address the lack of eye-tracking data in X-ray images by using desktop devices. However, these devices pose several concerns in data quality and data collection (Figure \ref{fig:reading_room}):

\begin{figure}[!b]
     \resizebox{\columnwidth}{!} {
    \includegraphics{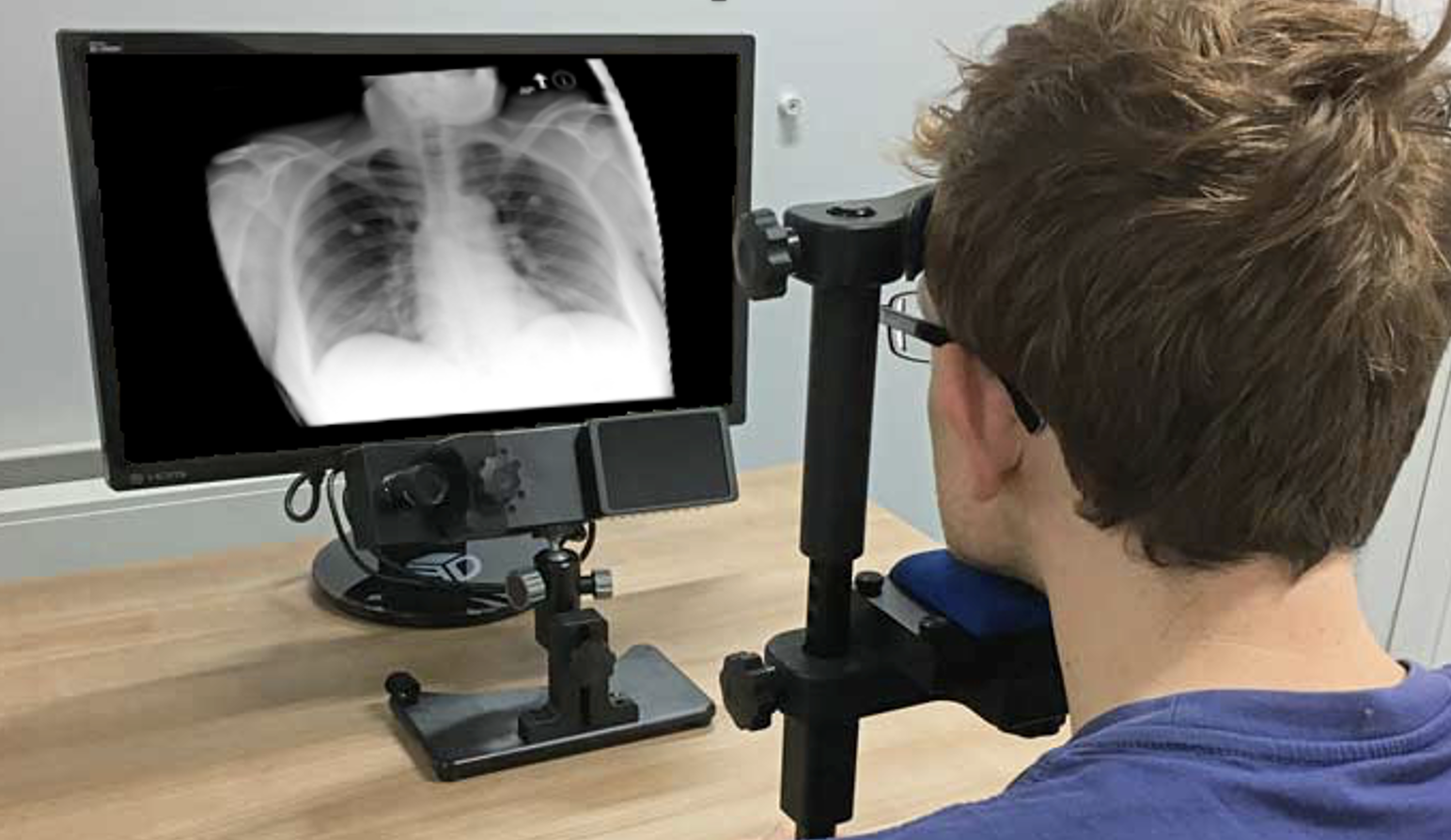}
    }
    \caption{EyeLink 1000 Plus \textregistered desktop eye tracker with chin and forehead support to avoid user doing lateral movements \small{(\url{https://www.sr-research.com/eyelink-1000-plus/}}). }
    \label{fig:reading_room}
\end{figure}

%\begin{itemize}[noitemsep,nolistsep]
    %\item 
    \noindent\textbf{Fixed distance from screen.} To collect eye-tracking data, radiologists need to be in a fixed position with a specific distance from the reading screen to ensure the system's calibration. This can %provoke the radiologists' 
    cause discomfort, increase %the 
    task difficulty, decrease concentration, and limit interaction with %the 
    X-ray images (e.g., zooming).
    %\vspace{0.1cm}
    
    % need to paraphrase
    %\item 
    \noindent\textbf{Frequent Calibration.} Eye-tracker calibrations are necessary for finding the correspondence between pupil and cornea positions in the image captured by the eye tracker’s camera and locations on the screen where the radiologists are looking \cite{Lanfredi2021reflacx}. %\\ %\vspace{-0.25cm}

    %\item 
    \noindent\textbf{Limited Head Movements and Loss of Line of Sight.} %Given that d
    Desktop eye-tracking devices require that the user is fixed to a particular position. Lateral body movements, %such as rotating the neck, 
    e.g. neck rotations %will lead to the loss of the eye-tracker 
    cause tracking loss and consequently lead to noise and data loss. 
    %\vspace{0.1cm}
    
    %\item 
    \noindent\textbf{Oculomotor drift may contribute to misinterpretations.} Measuring the eyes' drift and tremor with high precision and accuracy is challenging \cite{Niehorster21}. These movements appear to be beyond the resolution of the most available eye trackers.  %and they correspond to smooth gaze position signals that may be produced by noise suppression systems such as filters in the eye tracker’s software or hardware.
    %\vspace{0.1cm}
    
    %\item 
    \noindent\textbf{Difficulty of Interpreting Eye Tracking Data.} Eye-tracking analysis is based on the assumption that there is a relationship between fixation points and what the radiologist is thinking about. However, that is often not the case, and fixations do not necessarily translate into a conscious cognitive process. 
    \vspace{0.1cm}
%\end{itemize}

\vspace{-0.3cm}
\subsection{Using VR technologies for clinical practice}

In this paper, we argue that VR can 
overcome most of the desktop eye-tracking limitations %highlighted in traditional desktop eye-trackers and serve as a 
and become a disruptive technology for clinical practice in assessing X-ray images. %A VR headset is not bound to a fixed place, 
VR headsets enable %the 
radiologists to move comfortably and good ergonomic
postures when diagnosing, provide high-resolution screens that help radiologists analyze X-ray images, and can extract high-quality eye-tracking data and pupil dilations that AI frameworks can use to learn human classification patterns or that can be explored in HCI to infer correlations between eye movements and cognitive load.
The following sections discuss how this could be achieved based on recent advances. 

\section{Related Work}

Eye-tracking has been used extensively in HCI-related domains and is routinely applied to assess the impact of interface and web page designs on usability and content presentation. However, its application to improving human performance is still incipient. Additionally, eye movement data are notoriously difficult to analyze and heavily task-dependent. 
Duchowski's comprehensive book provides an excellent perspective of the field and its evolution \cite{Duchowski17book}. Recent surveys proposed measuring pupillary activity index as a base to models and estimate cognitive load\cite{Duchowski18chi,Wang21}. Other authors have looked at how perceived task difficulty in mental arithmetic affects microsaccadic rates, and magnitudes \cite{Siegenthaler14}. Older studies looked at using task-induced pupil diameter and blink rate to infer cognitive load \cite{Chen14pupil}. Kretjz et al. attempted to separate Ambient illumination, and Focal Attention \cite{Krejtz16}, whereas \cite{Helmert05} looked at Evidence from Static and Dynamic Scene Perception and its influence on eye movements. Minamoto et al. \cite{Minamoto14} studied how memory load affected eye movement control using the reading span test. %Older studies looked at eye fixations during text reading tasks to analyze how humans read text \cite{Just80}, \cite{Fitts50}.
Other authors have looked at Gaze Analytic methods to understand human behavior. More recently, \cite{mall2018modeling, khosravan2019,stember2019eye, aresta2020} have used eye-tracking data to improve segmentation and disease classification in Computed Tomography (CT) by integrating them in deep learning techniques. We argue that eye-movement data from trained radiologists can offer deep insights into how they analyze images and formulate diagnostics \cite{Gehrer18}.

\section{Advantages of VR Usage in Radiology Rooms}
A recent systematic literature review conducted by Trevia et al. \cite{Trevia21} analyzed the potential of VR technologies in radiology rooms to assist clinicians in their practice. The authors found some important points where VR technologies could be advantageous in clinical radiologist practices: 
%\vspace{0.1cm}

%\begin{itemize}[noitemsep,nolistsep,noindent]
    %% need to paraphrase 
    %\item[] %%%% Ganha-se algum espaço assim. Estamos a ficar apertados
    \noindent
    \textbf{Reduced Lightning and Ergonomics Equipment Costs}. The use of VR in radiology cuts the equipment and maintenance costs of a traditional radiology reading room, and it could potentially improve accuracy in radiological diagnosis \cite{Elsayed20}. 
    %\vspace{0.1cm}
    
    %\item[] 
    %\noindent
    %\textbf{Alternative to 3D printed models.} VR can provide a more flexible and inexpensive alternative of reproducing 3D printed models of patient-specific anatomical regions \cite{Venson17,Venson2016MedicalVR}.
    %\vspace{0.1cm}
    
    %\item[] 
    \noindent
    \textbf{VR Renders High Quality Data Visualisations}. This can enable more detailed surgery planning and communications among expert radiologists and patients in medical appointments. \cite{Venson17,Venson2016MedicalVR}.
    %\vspace{0.1cm}
    
    %\item[] 
    \noindent
    \textbf{VR can Promote Medical Training}. Students and experienced radiologists can leverage VR technologies for medical training, which can be reviewed at the learners' pace \cite{Knodel18}. This can provide a cost-effective alternative to face-to-face training sessions or simulations.
    %\vspace{0.1cm}

%\end{itemize}

To complement the above list, we add the following points, which we argue are key missing ingredients that provide advantages in the adoption of VR technologies for clinical practice:

%\begin{itemize}[noitemsep,nolistsep]
    %\item[] 
    \noindent\textbf{Accurate in Eye-Tracking Mechanisms.} VR technologies can overcome the limitations of current desktop eye-tracking devices and enable the collection of fixation points and eye gaze without the problem of losing line of sight or having body movements.
    %\vspace{0.1cm}
    
    %\item[] 
    \noindent\textbf{Insights about Radiologists Cognitive Load}. The duration of fixation points, the length of the saccades, pupil dilation, or even the number of blinks that a radiologist makes during a reading can provide insights into how complex the diagnosis is.
    
    %\item[] 
    \noindent\textbf{Insights about Radiologists' Fatigue}. A significant amount of literature suggests correlations between fixation points and pupil dilations with how tired a radiologist is \cite{Duchowski18chi}. With the high expectations in terms of pressure and speed in delivering a diagnosis, radiologists are affected by fatigue, which could be detected with the state of the art metrics and could alert the radiologist to take a break. 
    
    %\item[] 
    \noindent\textbf{Promotes the Development of Multi-Modal Learning AI.} Eye-tracking data and cognitive load features could be incorporated in advanced deep learning architectures to learn human classification patterns instead of focusing solely on the pixels of X-ray images. 
   %\vspace{0.1cm}
    
    %\item[] 
    \noindent\textbf{Novel Human-Centred XAI Mechanisms}. Current explainable AI algorithms fail to accurately identify the regions in x-ray images that are responsible for a given prediction. Incorporating eye-tracking data %has the potential of teaching the machine 
    can show how radiologists learn and how they assess %an 
    x-ray images. This could provide human-centric explanations to expert radiologists and promote trust in AI-based technologies.
   %\vspace{0.1cm}
    
    %\item[] 
    %\noindent
    \noindent\textbf{Simulations of healthcare environments for clinical practice}. VR technologies can use AI mechanisms to teach young student radiologists how to read an X-ray image based on the classification patterns of more senior radiologists. 
    %\vspace{0.1cm}
    
    \noindent
    \textbf{VR Promotes Remote Collaboration.} VR can also support joint work as it creates many opportunities for collaboration. In situations like pandemics where traveling needs to be minimized, VR technologies can bring together world radiologist experts to analyze and assess complex X-ray images in collaboration in a virtual reading room. This is different from tele-medicine or tele-diagnostics which are already feasible today. Indeed, real-time collaboration via shared virtual worlds can provide a fertile ground for research.

%In the following sections, we will focus 
In what follows, we discuss how eye-tracking can be combined with AI technologies to support, amplify, augment and enhance expert radiologists' performance when assessing X-ray images. 

\section{Eye-Tracking and AI for X-Ray Imagiology}

\begin{figure}[!b]
    \resizebox{\columnwidth}{!} {
    \includegraphics{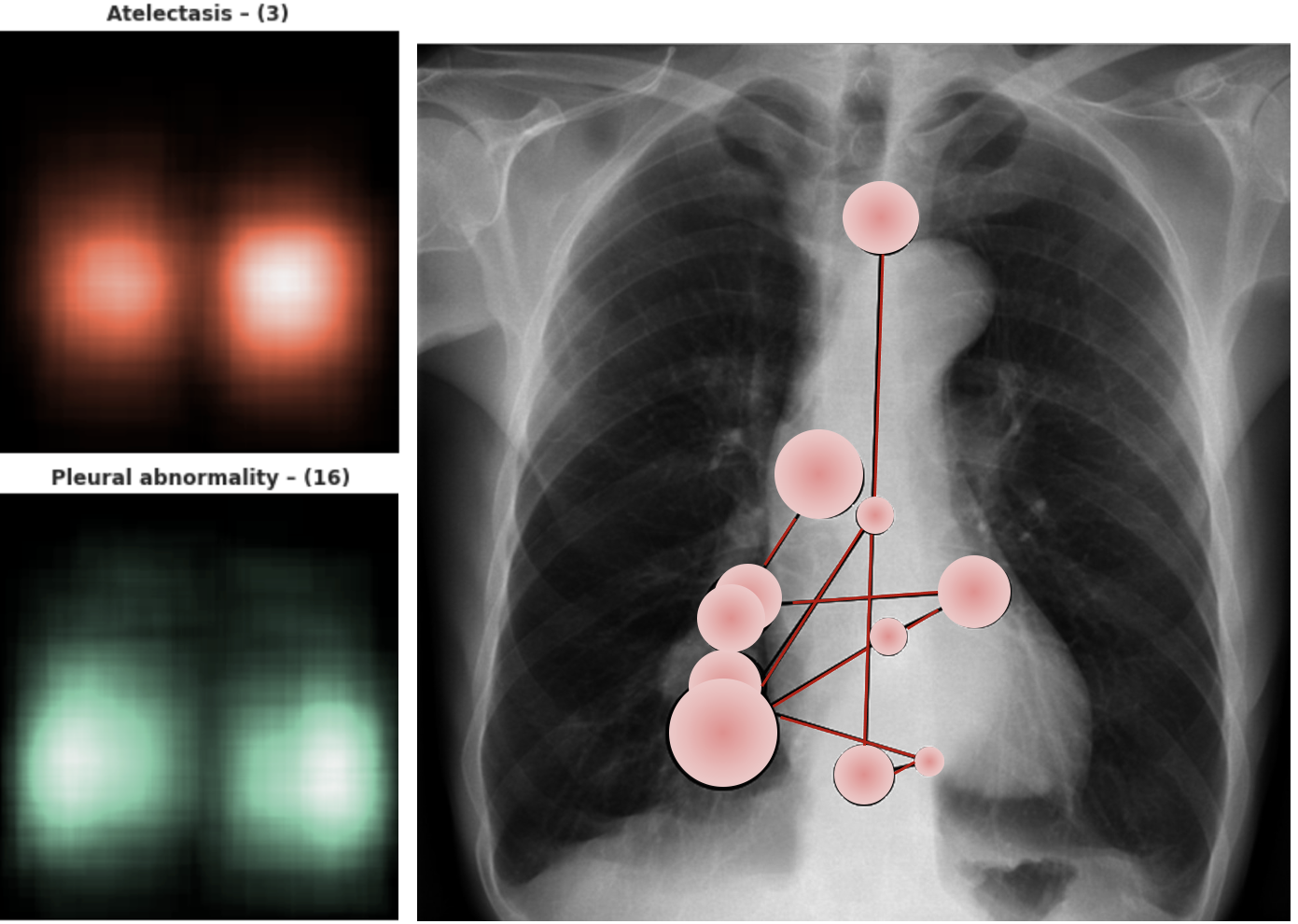}
    }
    \caption{Example of heatmaps generated from radiologists annotations for different lesions (left). Radiologist eye-tracking scan paths and fixation points (right) \cite{Lanfredi2021reflacx}.  }
    \label{fig:xray}
\end{figure}

The reduced cost of VR equipment promoted the adoption of eye-tracking technologies and the creation of novel X-ray datasets with multi-modal data such as fixation points, pupil dilations, radiologists audio utterances, etc \cite{Lanfredi2021reflacx,Karagyris21science}. Figure \ref{fig:xray} shows an example of heatmaps generated from radiologists' manual annotations in chest X-rays and an example of a radiologist's scan-path and fixation points generated from eye-tracking data \cite{Lanfredi2021reflacx}.

%Currently, some works in the literature combined different data modalities in deep learning architectures to achieve higher accuracy in predicting pneumonia in chest X-ray images \cite{Gale19reports}. For instance, Moon et al. \cite{moon2021multimodal} combined chest X-ray images and radiologists text reports to predict the disease of the patient and to automatically generate radiologists reports. However, these architectures are opaque and do not provide any understandings to the expert radiologist of the algorithms internal workings. This uncertainty make radiologists skeptic and resistant to the adoption of AI-based technologies in X-ray diagnosis. There is a need for  human-centered design aspects of explanations that must be human-understandable and domain-specific in AI-based predictions for X-ray images \cite{Chromik2021}. There is a growing need to design interactive explainable models that allow radiologists to drill down or ask for different explanations until (s)he is satisfied.

Currently, some works in the literature combined different data modalities in deep learning architectures to achieve higher accuracy in predicting pneumonia in chest X-ray images \cite{Gale19reports}. For instance, Moon et al. \cite{moon2021multimodal} combined chest X-ray images and radiologists' text reports to predict the patient's disease and to generate diagnostics automatically. However, these architectures are opaque and do not explain the algorithms' internal workings to the expert radiologist. This uncertainty makes radiologists skeptical and resistant to adopting AI-based technologies in X-ray diagnosis. Therefore, there is a need for human-centered explanations that are both human-understandable and domain-specific in AI-based predictions for X-ray images \cite{Chromik2021}. In addition, there is a growing need to design interactive explainable models that allow radiologists to drill down or ask for different explanations until they are satisfied.

A thought experiment carried out in 2018 at the Neural Information Processing Systems Conference asked hundreds of participants to imagine that they had a cancerous tumor and needed to choose a surgeon to operate them. The first option was a human surgeon who could explain the details of the operation but had a 15\% chance of causing death during the surgery. The second option was a robot that could achieve a 98\% success rate but not provide any explanations or answers to any questions. In the end, the robot surgeon received just one vote \cite{Rudin2019Why}. This thought experiment demonstrates the skepticism amongst people in accepting AI-based technologies and the need for human-centred AI technologies, particularly in contexts of high-stakes decision-making such as healthcare \cite{Rudin19}.

%Current AI tools make correct predictions in diagnosis, but not for the right reasons. Saliency maps which are used to generate explanations do not provide accurate medical diagnosis when compared with human radiologists. There is a growing need of novel techniques to make explanations accurate and consequently promote trust in the adoption of AI-based technologies. 

\subsection{Eye-Tracking as a Foundation for Explainable Human-Centric AI}
Very few studies in the literature have attempted to use eye-tracking radiologist data combined with chest X-ray images. We argue that eye-tracking is a key missing ingredient that can make machines learn how radiologists learn and serve as a foundation for developing human-centric AI technologies that can learn from human classification patterns. Eye-tracking can provide unique insights into AI learning architectures of how confident (and tired) the radiologist is in his readings, the degree of expertise, and the task complexity. 

Eye-tracking and pupil dilation data can help radiologists better understand the difficulty of assessing a specific X-ray image. According to Duchowski et al. \cite{Duchowski18chi} fixation duration is proportional to the duration of the underlying difficulty of the cognitive operation, longer fixations and shorter saccades may suggest increased cognitive load, and more complex problems evoke large pupillary dilations. Additionally, eye-tracking scan paths can also indicate the degree of expertise of the radiologist: targeted fixation points correlate with expertise. Studies suggest that the more expert a radiologist is, the more targeted the eye-tracking scan paths are. In Borys and Plechawska-W\'{o}jcik \cite{Borys2017}, the authors asked a group of students and an expert radiologist to read a chest X-ray image. They found that the scan paths of the expert radiologist were sharp and targeted at the regions of the X-ray with lesions. 
Students who could read the image correctly had messier scan paths with larger fixation points in the regions with lesions. Those who could not read the image had random scan paths and shorter duration fixation points.

\section{Conclusions}

A VR headset provides high-resolution screens that help radiologists analyze X-ray images and extract high-quality eye-tracking data and pupil dilations. AI frameworks can use this data to learn human classification patterns and discover correlations between eye movements and cognitive load. In this paper, we provide a discussion of the advantages of VR equipment in the collection and generation of eye-tracking data to support, amplify, augment and enhance expert radiologists' performance when assessing X-ray images. We put forward two main arguments: (1) eye-tracking data has the potential to enable novel AI learning architectures that, instead of learning solely from X-ray pixel data, will learn from radiologists' classification patterns that are encoded in their eye movements; and (2) VR can overcome most of the limitations highlighted in traditional desktop eye-trackers and serve as a disruptive technology for clinical practice in assessing X-ray images.

We finish %this paper 
by reflecting on major challenges affecting the adoption of VR technologies in clinical practice. The first is device physical characteristics and resolution. While radiologists complained about the display resolution of earlier VR head-mounted displays (HMDs), have progressed at a faster clip. They are now competitive in terms of pixels/radian with conventional displays. We expect time to tilt the scales even more in favor of HMDs. While over ten million Oculus Quest 2 units were shipped in 2021, Reading Room displays remain an expensive niche market.
In contrast, commodity HMDs now cost comparatively much less. %While we argue 
As for eye-tracking interfaces, the required hardware is not part of commodity headsets. %such as the Oculus 2. 
However, this is changing. Many commercial headsets now feature infrared cameras for this purpose, at a twentieth of the cost of desktop eye-tracking a decade ago. Finally, the most significant barrier to the adoption of VR HMDs as medical devices has to do with regulatory approval. Indeed, while Computer Science research requires comparatively small user studies, establishing the clinical validity of a novel medical device or procedure requires extensive testing and validation to warrant regulatory approval. Still, given the many advantages and promises outlined herein, we believe that the future of clinical radiology diagnostics lies in Extended Reality Technologies.

\acknowledgments{
This work was supported in part by the Portuguese Govt. \textit{Funda\c{c}\~{a}o para a Ci\^{e}ncia e a Tecnologia}, under project UIDB/50021/2020.}

%\bibliographystyle{abbrv}
%\bibstyle{unsrt2authabbrvpp}
%\bibliographystyle{abbrv-doi}
%\bibliographystyle{abbrv-doi-narrow}
%\bibliographystyle{abbrv-doi-hyperref}
\bibliographystyle{abbrv-doi-hyperref-narrow}

%\bibliography{template, my_works}

\end{document}